\documentclass[twocolumn,preprintnumbers,amsmath,amssymb,superscriptaddress,showpacs]{revtex4}
\usepackage{epsfig}
\usepackage{graphicx}

\begin{document}
\title{Asymmetric phase-covariant {\em{d}}-dimensional cloning}

\author{L.-P. Lamoureux}
\affiliation{Quantum Information and Communication, Ecole
Polytechnique, CP 165, Universit\'e Libre de Bruxelles, 1050
Brussels, Belgium}

\author{N. J. Cerf}
\affiliation{Quantum Information and Communication, Ecole
Polytechnique, CP 165, Universit\'e Libre de Bruxelles, 1050
Brussels, Belgium}

\begin{abstract}
We consider cloning transformations of {\em{d}}-dimensional states
of the form  $e^{i\phi_0}|0\rangle + e^{i\phi_1}|1\rangle +\dotso +  e^{i\phi_{d-1}}|d-1\rangle$ 
that are covariant with respect to rotations of the phases $\phi_i$'s.
The optimal cloning maps are easily obtained using a well-defined general characterization of state-dependent $1 \rightarrow 2$ cloning transformations
in arbitrary dimensions. Our results apply to symmetric as well as asymmetric
cloners, so that the balance between the fidelity of the two clones can be analyzed.
\end{abstract}
\maketitle

\section{INTRODUCTION}
Quantum mechanics prohibits the perfect cloning of an unknown quantum state \cite{WZ}.  This discovery led way to new and provably secure protocols for quantum key distribution (QKD) \cite{BB84,Ekert}.  Indeed, one of the main ingredients of the security of QKD is the impossibility of perfectly cloning unknown quantum states selected from a nonorthogonal set. As a result, a potential evesdropper Eve can only make approximate copies of the states emitted by a sender Alice \cite{BH,SM,Brussetal,Werner,Paulicloner}.  She does so by optimizing the fidelity of the copies with respect to the original state.  She then keeps one to herself and sends the other copy to a receiver Bob.  Such an evesdropping strategy has been shown to be optimal for many QKD protocols \cite{BM,CBKG}.  Apart from cryptography, quantum cloning and the underlying sharing of quantum information between several parties is an interesting feature to study in itself. \par  

A typical feature of quantum cloning is that the optimal cloning transformation depends on the considered set of input states.  The greater the set, the lower the fidelity. More precisely, for group covariant cloning \cite{DL} where the set of input states is the orbit of a given state under the action of a group of unitary transformations the smaller is the group, the higher is the fidelity averaged over the input states.
\par   

In this paper, we analyze quantum cloning machines (QCMs) that duplicate
with an equal fidelity all uniform $d$-dimensional superposition states
with arbitrary phases. These so-called phase-covariant cloners have
been found initially for qubits ($d=2$) \cite{BCAM,CDG} 
and qutrits ($d=3$) \cite{DL,CDG,AM}, and we shall investigate their
$d$-dimensional extension here. Note that optimal $N\rightarrow M$ phase-covariant cloning maps also have been derived analytically for higher dimensional systems in \cite{buscemi} although it was only for special values of $M$.  Here, we use a convenient cloning formalism
which is based on the idea that, after the cloning transformation, the clones are left in a mixture of the input state itself 
and states resulting from applying operators of the discrete Weyl group 
(also called "error" operators) on the input state \cite{cerf1,cerf2}. 
As we shall see, our approach will automatically lead to the optimal covariant cloning under the Abelian group $U(1)^{\otimes (d-1)}$ of phase rotations -- 
the phase-covariant {\em{d}}-dimensional cloning. 
Although we give no analytical proof that our transformation is optimal, 
it has been shown very recently to be the case for symmetric cloners
by Fan {\em et al.} in \cite{FIMW}.  In their paper, Fan {\em et al.} calculate the optimal $1 \rightarrow 2$ cloning map for the set of $d$-dimensional input states $e^{i\phi_0}|0\rangle + e^{i\phi_1}|1\rangle +\dotso +  e^{i\phi_{d-1}}|d-1\rangle$ by considering the most general cloning transformation where the only assumption made is that the cloner is symmetric, {\em i.e.}, one has identical cloning maps for both clones.  The drawback is that the calculation of the reduced density matrices of the clones and maximization of the fidelity is somewhat tedious.  Our method allows to recover the same result in a much simpler way, as well as extending it very naturally
to the asymmetric case. In the latter case, we have checked the optimality
of our construction by numerically computing the best asymmetric cloning transformation (for several asymmetry parameters),
and verifying that the results coincide up to the machine precision.   
\par 

The paper is organized as follows.  In Sec. II, we describe a fairly general class of (state-dependent) quantum cloning machines \cite{cerf1,cerf2}.  
We then apply this formalism to the case of $d$-dimensional phase-covariant cloning and obtain the optimal value of the fidelity in the symmetric case.
In Sec. III, we extend our calculations to the case of asymmetric cloning. 
Finally, in Sec. IV, we conclude with a comparison of the performance 
of the optimal $d$-dimensional phase covariant cloner 
with the other known cloners in $d$ dimensions, namely the universal cloner \cite{Werner,cerf1,BH2}, the cloner of real states \cite{Navez}, 
and the cloner of two mutually unbiased bases \cite{CBKG}.

\section{OPTIMAL PHASE-COVARIANT CLONING}

We use a general class of cloning transformations 
as defined in Refs. \cite{cerf1, cerf2}.
Consider an arbitrary state $|\psi\rangle$ in a $d$-dimensional Hilbert space of which we wish to produce two (approximate) clones.  The class of cloning transformations we will analyze is such that, if the input state is $|\psi\rangle$, then the resulting joint state of the two output clones (noted $A$ and $B$) and the cloning machine (noted $C$) is:
\begin{eqnarray}
|\psi\rangle&\rightarrow& \sum^{d-1}_{m,n=0}a_{m,n}U_{m,n}|\psi\rangle_A|B_{m,-n}\rangle_{B,C}\nonumber\\
&=&\sum^{d-1}_{m,n=0}b_{m,n}U_{m,n}|\psi\rangle_B|B_{m,-n}\rangle_{A,C}
\end{eqnarray}
where
\begin{equation}
U_{m,n} = \sum^{d-1}_{k=0}e^{2i\pi(kn/d)}|k+m\rangle\langle k|
\end{equation}
and
\begin{equation}
|B_{m,n}\rangle = \frac{1}{\sqrt{d}}\sum^{d-1}_{k=0}e^{2i\pi(kn/d)}|k\rangle|k+m\rangle
\label{bell}
\end{equation}
with $0\le m,n \le d-1$. Here, $U_{m,n}$ is an element of the discrete
Weyl group or a so-called error operator : it shifts the state by $m$ units (modulo $d$) in the computational basis, and multiplies it by a phase so as to shift its Fourier transform by $n$ units (modulo $d$). Equation (\ref{bell}) defines the $d^2$ generalized Bell states for a pair of $d$-dimensional systems.
\par

Tracing over systems $B$ and $C$ (or $A$ and $C$) yields the final state of clone $A$ (or clone $B$):  if the input state is $|\psi\rangle$, then the clones $A$ and $B$ are left in a mixture of the states $|\psi_{m,n}\rangle = U_{m,n}|\psi\rangle$ with respective weights $|a_{m,n}|^2$ and $|b_{m,n}|^2$:
\begin{eqnarray}
\rho_A &=& \sum_{m,n=0}^{d-1}|a_{m,n}|^2\,|\psi_{m,n}\rangle\langle\psi_{m,n}|\nonumber\\
\rho_B &=& \sum_{m,n=0}^{d-1}|b_{m,n}|^2\,|\psi_{m,n}\rangle\langle\psi_{m,n}|
\label{output}
\end{eqnarray}
In addition, the weights of the two clones are related through the fact that
the amplitude functions $a_{m,n}$ and $b_{m,n}$ 
are dual under a Fourier transform \cite{cerf1, cerf2}:
\begin{equation}
b_{m,n} = \frac{1}{d}\sum^{d-1}_{x,y=0}e^{2i\pi(nx-my)/d}a_{x,y}.
\label{ft}
\end{equation}
The fidelity of a clone, say $A$, is given by
\begin{equation}
F_A = \langle \psi |\rho_A| \psi \rangle = \sum^{d-1}_{m,n=0}|a_{m,n}|^2|\langle \psi |\psi_{m,n}\rangle|^2
\label{fidelity}
\end{equation}
and similarly for the clone $B$ 
(replace the $|a_{m,n}|^2$ term by $|b_{m,n}|^2$).
\par

Let us now analyze the possibility of using this general formalism in order
to clone the class of $d$-dimensional states that can be expressed as
\begin{equation}
|\psi\rangle = \frac{1}{\sqrt{d}}\sum^{d-1}_{k=0}e^{i\phi_k} \, |k\rangle.
\label{psiphi}
\end{equation}
Inserting this state $|\psi\rangle$ 
in Eq.~(\ref{fidelity}) yields the expression for the fidelity $F_A$
with
\begin{eqnarray}
\langle \psi|\psi_{m,n} \rangle &=& \frac{1}{d}
\sum^{d-1}_{k=0}e^{i(\phi_k-\phi_{k+m})} \,
e^{2i\pi (nk)/d}.
\label{fidphi}
\end{eqnarray}
Note that if $m=0$, then the identity $\sum^{d-1}_{k=0} e^{2i\pi(nk)/d} = d\,\delta_{n,0}$ implies that
$\langle \psi|\psi_{m,n} \rangle = \delta_{n,0}$, so that all the elements of the $a_{m,n}$ matrix with $m=0$ and $n\ne 0$
do not contribute to the fidelity $F_A$.
We are interested in a cloning machine that yields the same fidelity
for all possible values of the $\phi_j$'s.  This imposes strong contraints on the amplitudes $a_{m,n}$ characterizing the cloner.  A form which does satisfy these constraints is expressed by the following amplitude matrix
\begin{eqnarray}
a_{m,n} =
\left ( \begin{array}{cccc}
v & y &\cdots  & y\\
x & x &\cdots  & x \\
\vdots &  &  & \vdots \\
x & x & \cdots & x
\end{array} \right)
\end{eqnarray}
where $v$, $x$, and $y$ are arbitrary constants
satisfying the normalization constraint
\begin{equation}
v^2+(d-1)y ^2+d(d-1)x^2 = 1.
\label{normalization}
\end{equation}
Replacing these $a_{m,n}$ values in Eq.~(\ref{fidphi}) yields
for the fidelity
\begin{equation}
%F_A &=& v^2 + \sum^{d-1}_{m=1}\sum^{d-1}_{n=0} x^2 \cdot %\frac{1}{d^2}\sum^{d-1}_{k,k'=0}
%e^{i(\phi_k-\phi_{k+m})+2i\pi\frac{kn}{d}}e^{i(\phi_{k'+m}-\phi_{k'})-2i\pi\frac{k'n}{d}%}\nonumber\\
F_A = v^2 + (d-1)x^2
\end{equation}
which is indeed independent of $y$. Note that for a universal cloner,
one has $x=y$ \cite{cerf1,cerf2}. The main idea here is that we can have 
$x>y$, implying that the cloning fidelity for the states (\ref{psiphi})
can be higher than that of the universal cloner,
at the expense of a lower fidelity for the states of
the computational basis $\{|k\rangle\}$.
\par

The second clone is characterized by a similar amplitude matrix
\begin{eqnarray}
b_{m,n} =
\left ( \begin{array}{cccc}
v' & y' &\cdots  & y'\\
x' & x' &\cdots  & x' \\
\vdots &  &  & \vdots \\
x' & x' & \cdots & x'
\end{array} \right).
\end{eqnarray}
where the different matrix elements are related to the $a_{m,n}$ coefficients by Eq.~(\ref{ft}):
\begin{eqnarray}
v' &=& \frac{1}{d}[v+(d-1)y+d(d-1)x]\\
y' &=& \frac{1}{d}[v+(d-1)y-dx]\\
x' &=& \frac{1}{d}(v-y).
\end{eqnarray}
Since we seek a symmetric cloner, the amplitude coefficients for the two clones must be equal, that is
\begin{eqnarray}
v &=& v' = \frac{V+(d-1)X}{\sqrt{d}}\\
y &=& y' = \frac{V-X}{\sqrt{d}}\\
x &=& x' = \frac{X}{\sqrt{d}}.
\end{eqnarray}
where we have introduced two new parameters $V$ and $X$ (the third parameter
has been eliminated by the symmetry constraint). In addition,
as a result of the normalization condition
\begin{equation}
V^2+2(d-1)X^2=1 ,
\label{norm}
\end{equation}
only one free parameter remains. In this new parametrization, 
the expression of the fidelity (for both clones) reduces to
\begin{equation}
F = \frac{1+(d-1)(d-2)X^2+2(d-1)VX}{d}.
\label{fidfinal}
\end{equation}
We are interested in finding the values of $V$ and $X$ that maximize Eq.~(\ref{fidfinal}) under the normalization constraint (\ref{norm}).
A simple maximization by use of Lagrange multipliers yields
\begin{equation}
F_{opt} = \frac{1}{d}\bigg[1+\frac{d-2+\sqrt{(d-2)^2+8(d-1)}}{4}\bigg],
\label{fopt}
\end{equation}
which coincides with the fidelity of the
optimal $1\rightarrow 2$ phase-covariant QCM for any dimension 
as found in \cite{FIMW}.  The corresponding solutions of $V$ and $X$ are
\begin{eqnarray}
V^2 &=& \frac{d-1}{d}\frac{1}{1+(d-2)F_{opt}}\\
X^2 &=& \frac{1}{2(d-1)} - \frac{1}{2d}\frac{1}{1+(d-2)F_{opt}}.
\end{eqnarray}
\par

\begin{figure}[tb!]
\centerline{\includegraphics[width=0.9\linewidth]{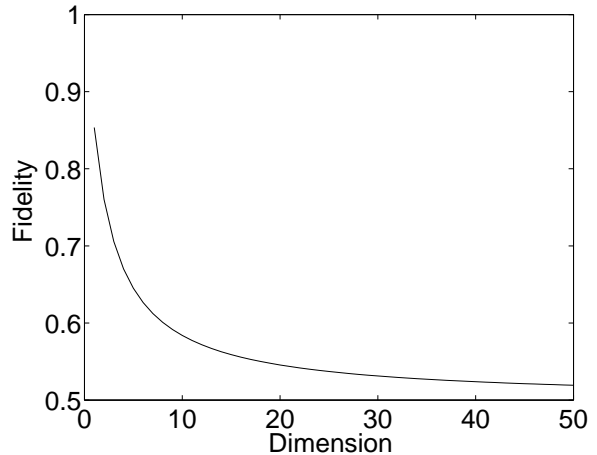}}
\caption{Fidelity of the clones versus the dimension for the symmetric phase-covariant cloner.  The continuous line is there for visual purposes only.} \label{fig1}
\end{figure}

In Figure 1, we plot the fidelity of the clones as a function of the dimension.
Note that the cloning fidelity $F_{opt}$ tends to 1/2 + $O(1/d)$
in the high-dimensional limit. Indeed, in this limit, the cloner operates
simply by moving the input state into one of the clones, chosen at random,
while preparing the other clone in the fully mixed state.  As the latter has a vanishing contribution to the fidelity in the high-dimensional limit, we get a fidelity of $\frac{1}{2}$.
In the special case where $d=2$, we recover the optimal 
phase-covariant qubit cloner of fidelity \cite{BCAM,CDG}
\begin{equation}
F_{d=2}=\frac{1}{2}(1+1/\sqrt{2})\simeq 0.854
\end{equation}
while, when $d=3$, we recover the optimal two-phase-covariant qutrit cloner \cite{DL,CDG}
\begin{equation}
F_{d=3}=\frac{5+\sqrt{17}}{12}\simeq 0.760
\end{equation}

\section{ASYMMETRIC CLONING}

We now generalize Eq.~(\ref{fopt}) to asymmetric cloning transformations ($F_A \neq F_B$).  The cloning fidelity of the first clone can be expressed 
as a function of the fidelity of the second clone (again using the same forms
of the $a_{m,n}$ and $b_{m,n}$ matrices) as
\begin{eqnarray}  
F_A(F_B,v,y) =  {v}^{2}
+ \Delta
\label{fidas}
\end{eqnarray}
and the normalization as 
\begin{eqnarray}  
 v^2 + (d-1)y^2 + \Delta =  1
\end{eqnarray}
where
\begin{widetext}
\begin{equation}
\Delta = {\frac {\left (-2yd-2v+2y+2\sqrt {2ydv-{y}^{2}d
+{v}^{2}-2vy+{y}^{2}+{\it F_B}{d}^{2}-d{v}^{2}}\right )^{2}}{
 4(d-1){d}^{2}}}.
\end{equation}
\end{widetext}
We then maximize $F_A$ over $v$ and $y$ while keeping $d$ and $F_B$ constant,
yielding the best possible balance between the fidelities $F_A$ and $F_B$.

Figure 2 displays the resulting shrinking factor $\eta_A$ of the first clone as a function of the shrinking factor of the second clone $\eta_B$.  The shrinking factor $\eta$ is defined here as the probability for the input state not to be depolarized, that is, $F=\eta +(1-\eta)/d$.
%related to the fidelity by $\eta = \frac{d\ F - 1}{d-1}$ for a given clone and dimension.  
We see that the fidelity of a clone is equal to one (or $\eta=1$) 
when the fidelity of the second clone is equal to $1/d$ (or $\eta=0$),
i.e., when it is completely mixed.  We also confirm that the quality of the clones diminishes as a function of the dimension.

\begin{figure}[htb]
\centerline{\includegraphics[width=0.9\linewidth]{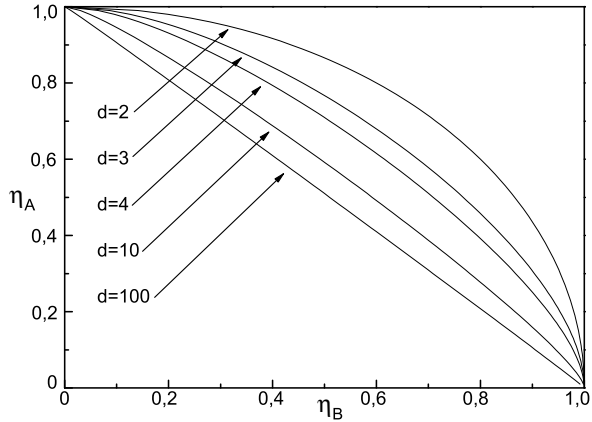}}
\caption{Shrinking factor of clone $A$ as function of the shrinking factor of clone $B$ for $d=2$,3,4,10 and 100.} \label{fig2}
\end{figure}

We have numerically confirmed the optimality of this class of
asymmetric (and therefore symmetric) cloners with the use of a technique based on semidefinite programming \cite{Fiurasek01, Fiurasek03}. 
The cloning transformation is a linear trace-preserving completely positive (CP) map that can be represented by a positive semidefinite operator $S$ on the tensor-product space of the input and output states.  The fidelities can be expressed as $F_{A(B)} = Tr[SR_{A(B)}]$
with an appropriately defined operator $R_{A(B)} \ge 0$ \cite{Fiurasek03}.
 The optimal asymmetric cloner can be obtained by maximizing $F = pF_A + (1-p)F_B$, where $p \in [0,1]$ is an asymmetry parameter.  The resulting fidelities coincide with those obtained when maximizing Eq.~(\ref{fidas}) 
up to the machine precision.

\section{CONCLUSION}

We have found the class of optimal $1\rightarrow 2$ phase-covariant QCMs 
in any dimension $d$ along with their corresponding fidelities,
Eq.~(\ref{fopt}).
Although it was not demonstrated analytically but only checked numerically,
this cloner has been proved to be optimal in \cite{FIMW} in the symmetric
case. Interestingly, almost all cloners that have been considered on the basis of the formalism of \cite{cerf1,cerf2} have, so far, always yielded 
the optimal fidelity when it could be compared to another calculation  \footnote[23]{The only exception found to date concerns the 2-state cloner introduced in connection with the B92 protocol \cite{Brussetal}, the reason being that the formalism of \cite{cerf1, cerf2} is restricted to unital (input-to-single-clone) maps.}.  
The $d$-dimensional symmetric phase-covariant cloner is just another
example of this.
In the special case where $d=2$ and $d=3$, we recover the optimal phase-covariant qubit and qutrit cloners.
Furthermore, we have extended our investigation to the class of asymmetric QCMs, and concluded that the relative fidelity between two clones decreases with the dimension.
\par

In Figure 3, we have plotted, for comparaison, the fidelity as a function of the dimension $d$ for the universal cloner \cite{Werner,cerf1,BH2},
the real cloner \cite{Navez}, the optimal cloner of two mutually unbiased bases \cite{CBKG}, and the optimal phase-covariant cloner.  
As expected, the universal $d$-dimensional cloner has a lower fidelity 
\begin{equation}
F_U = (d+3)/(2(d+1))
\end{equation}
than the other cloners since they span smaller sets of states. The cloner of two mutually unbiased bases is the one which spans the smaller set and therefore has the highest fidelity 
\begin{equation}
F_{MU} = \frac{1}{2} (1+\frac{1}{\sqrt{d}}).
\end{equation}
In between lies the real cloner 
\begin{equation}
F_R =\frac{1}{2}+ \frac{\sqrt{d^2+4d +20} -d +2}{4(d+2)}
\end{equation}
and phase-covariant cloner.  Except when $d=2$ where the fidelities 
of the latter two cloners are equal to that of the cloner of two mutually unbiased bases, the fidelity of the $d$-dimensional phase-covariant cloner remains close to the universal cloner and slightly below the real cloner.
\par

\begin{figure}[htb]
\centerline{\includegraphics[width=0.9\linewidth]{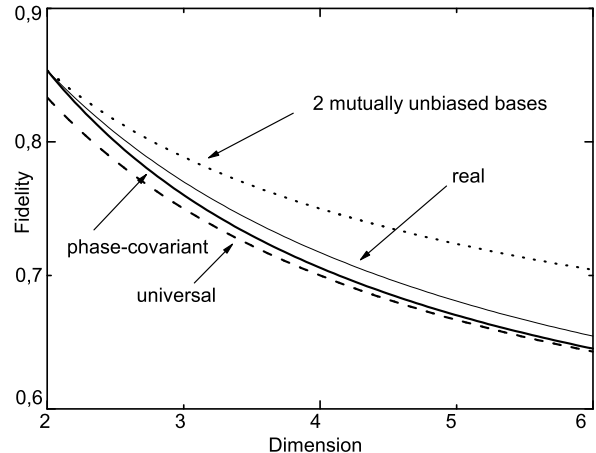}}
\caption{Fidelity $F$ as a function of the dimension $d$ for the universal cloner (dash) \cite{Werner,BH2}, the real cloner  \cite{Navez}, the phase-covariant cloner derived in this paper and the cloner of two mutually unbiased bases (dot) \cite{CBKG}.  The lines are there for visual purposes.} \label{fig3}
\end{figure}

\section{ACKNOWLEDGMENTS}
We thank T. Durt and C. Macchiavello for useful discussions.
We acknowledge financial support from the Communaut\'e Fran\c{c}aise
de Belgique, from the IUAP programme of the Belgian government and
from the EU under project RESQ in the IST/FET programme.

\end{document}